\documentclass[conference]{IEEEtran}
\IEEEoverridecommandlockouts

\usepackage{cite}
\usepackage{amsmath,amssymb,amsfonts}
\usepackage{algorithmic}
\usepackage{graphicx}
\usepackage{textcomp}
\usepackage{xcolor}
\def\BibTeX{{\rm B\kern-.05em{\sc i\kern-.025em b}\kern-.08em
    T\kern-.1667em\lower.7ex\hbox{E}\kern-.125emX}}

\usepackage{todonotes}
\usepackage{comment}
\usepackage[acronym]{glossaries}
\usepackage[hidelinks]{hyperref}

\newcommand{\Rom}[1]{\uppercase\expandafter{\romannumeral#1}}

\usepackage{cite}

\usepackage{tikz}
\usetikzlibrary{arrows.meta}
\usetikzlibrary{decorations.pathreplacing}
\usetikzlibrary{calc}

\usepackage{xcolor}
\usepackage{svg}
\usepackage{pgfplots}
\usepackage{adjustbox}
\usepackage{eso-pic}

\usepackage[caption=false,font=footnotesize]{subfig}

\newcommand\acceptedtext{%
  \footnotesize This work has been accepted for publication at the \par
  IEEE/ITSS Intelligent Transportation Systems Conference (ITSC), 2025, \textcopyright IEEE }
 
\newcommand\acceptednotice{%
  \AddToShipoutPictureBG*{%
    \begin{tikzpicture}[remember picture,overlay]
      \node[anchor=south,yshift=13pt] at (current page.south) {\parbox{0.65\textwidth}{\centering\acceptedtext}};
    \end{tikzpicture}%
  }%
}

\newcommand{\greycircle}{\tikz[baseline=-0.75ex]{\filldraw[fill=gray!60, draw=black] (0,0) circle (0.1cm);}}

\newcommand{\redzone}{\tikz[baseline=-0.75ex]{\filldraw[red!60!black, opacity=0.3] (0,0) circle (0.1cm);}}
\newcommand{\greenzone}{\tikz[baseline=-0.75ex]{\filldraw[green!60!black, opacity=0.3] (0,0) circle (0.1cm);}}

\graphicspath{{./img/}}

\begin{document}

\title{Rendezvous and Docking of Mobile Ground Robots for Efficient Transportation Systems
}
\author{\IEEEauthorblockN{1\textsuperscript{st} Lars Fischer}
\IEEEauthorblockA{
\textit{FZI Research Center}\\ \textit{for Information Technology}\\
Karlsruhe, Germany \\
lars.fischer@fzi.de}
\and
\IEEEauthorblockN{2\textsuperscript{nd} Daniel Flögel}
\IEEEauthorblockA{
\textit{FZI Research Center}\\ \textit{for Information Technology}\\
Karlsruhe, Germany \\
floegel@fzi.de}
\and
\IEEEauthorblockN{3\textsuperscript{rd} Sören Hohmann}
\IEEEauthorblockA{\textit{Institute of Control Systems} \\
\textit{Karlsruhe Institute of Technology}\\
Karlsruhe, Germany \\
soeren.hohmann@kit.edu}

}

\maketitle

\begin{abstract}
In-Motion physical coupling of multiple mobile ground robots has the potential to enable new applications like in-motion transfer that improves efficiency in handling and transferring goods, which tackles current challenges in logistics.
A key challenge lies in achieving reliable autonomous in-motion physical coupling of two mobile ground robots starting at any initial position.
Existing approaches neglect the modeling of the docking interface and the strategy for approaching it, resulting in uncontrolled collisions that make in-motion physical coupling either impossible or inefficient.
To address this challenge, we propose a central \acrfull{mpc} approach that explicitly models the dynamics and states of two omnidirectional wheeled robots, incorporates constraints related to their docking interface, and implements an approaching strategy for rendezvous and docking.
This novel approach enables omnidirectional wheeled robots with a docking interface to physically couple in motion regardless of their initial position.
In addition, it makes in-motion transfer possible, which is 19.75\% more time- and 21.04\% energy-efficient compared to a non-coupling approach in a logistic scenario.
\end{abstract}


\acceptednotice
\newacronym{agv}{AGV}{automated guided vehicle}
\newacronym{amr}{AMR}{autonomous mobile robots}
\newacronym{mpc}{MPC}{model predictive control}
\newacronym{dempc}{dMPC}{decentralized model predictive control}
\newacronym{dimpc}{DMPC}{distributed model predictive control}
\vspace{-1em}

\section{INTRODUCTION}
The importance of transportation is evident in its role in supporting all economic activities \cite{sivanandhamPlatooningSustainableFreight2020}.
In the production of goods, for example, material flow requires logistics to ensure the right product is in the right place at the right time \cite{bauernhanslIndustrie40Produktion2014}.
In logistics, \acrfull{amr} are often used for transportation due to increased efficiency compared to humans and thus, play a central role in modern logistic systems \cite{piknerCyberphysicalControlSystem2021}.
It is further postulated that the optimal logistic space is empty and only \acrshort{amr} swarms are used for transportation \cite{bauernhanslIndustrie40Produktion2014}.
However, this requires that \acrshort{amr} are able to handle a broad range of tasks in modern logistics \cite{heussModularRobotSoftware2019}.
While \acrshort{amr} still struggle with limited flexibility and reconfigurability in respond to changing needs, they are still seen as a key driver to gather flexibility and adaptability in transportation \cite{heussModularRobotSoftware2019}.
This imposes the need for \acrshort{amr} to be flexible, scalable, and adaptable \cite{sparcH2020_Robotics_MultiAnnual_Roadmap_ICT2017B2017,bauernhanslIndustrie40Produktion2014}.
Additionally, there is a need to efficiently transfer goods while driving to reduce occupancy time on loading and unloading areas \cite{sparcH2020_Robotics_MultiAnnual_Roadmap_ICT2017B2017}.

\begin{figure}[!t]
    \centering 
    \includegraphics[width=0.95\linewidth]{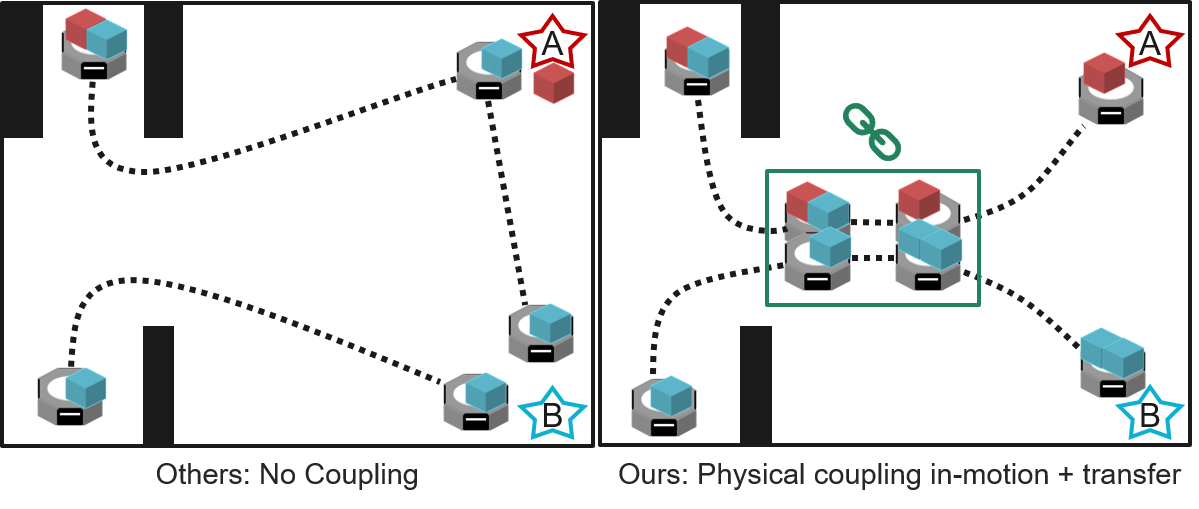}
    \caption{The proposed approach allows robots to physically couple in-motion and enables in-motion transfer. In a logistic scenario, this facilitates transporting packets with the same destination together in bundles, thereby reducing overall travel distance, energy consumption, and time.
    }
    \label{fig:paper_figure}
\end{figure}

Current motion planning approaches commonly implement collision avoidance by defining a safety distance between them \cite{ebelDistributedModelPredictive2017,stombergCooperativeDistributedMPC2023}.
However, it would be beneficial if they have the ability to physically couple in motion with each other.
First, this enables in-motion transfer, the ability to transfer goods while being in motion \cite{khanApplicationModularVehicle2023}.
Considering a logistic scenario as depicted in Fig.~\ref{fig:paper_figure}, the ability to physically couple enables robots to bundle packets with the same destination, which increases efficiency.
In the mobility domain, in-motion transfer can improve energy consumption, capacity utilization rate, and overall efficiency \cite{wuModularAdaptiveAutonomous2021,zhangModularTransitUsing2020,khanApplicationModularVehicle2023}.
Second, physically coupling a robot transporting a workpiece with another robot equipped with a tool enables the processing of workpieces during transportation.
Third, skill sets can be broadened, such as increasing maximum payload and loading capacity.
However, this requires robots with the ability to couple in motion physically.

Formation control methods that enable logical coupling are commonly used in logistics to implement cooperative tasks, e.g., cooperative object transportation, where multiple robots carry objects that are either too heavy, too large, or of a complex shape for one robot to transport \cite{ebelOptimizationDrivenControlOrganization2019,ebelDistributedModelPredictive2017,hompelTechnicalReportLoadRunner2020,tuciCooperativeObjectTransport2018}
However, these approaches do not model a docking interface or an approaching strategy, so setting the inter-robot distance to zero in these approaches results in uncontrolled collisions.
Besides, some approaches implement physical coupling for mobile ground robots, but either physical coupling is only possible if one agent is static \cite{hompelTechnicalReportLoadRunner2020,zhouDesignFabricationControl2022,zhouReconfigurableModularVehicle2022} or the feasibility of in-motion physical coupling is dependent on specific initial states \cite{yiConfigurationControlPhysical2022}, which is not efficient.

Therefore, there is a gap in the ability of mobile ground robots within the logistics domain to physically couple in motion, which limits their flexibility and adaptability by not enabling use cases like in-motion transfer.
To address these drawbacks, we propose a central \acrfull{mpc} motion planning approach, which includes the imposed constraints of the docking interface and the approaching strategy.

Therewith, it is achieved that \acrshort{amr} are able to physically couple in motion, regardless of their initial state.

The main contribution of this paper is a novel \acrshort{mpc} approach enabling mobile ground robots to physically couple in motion.
Therefore, we model and identify the key constraints necessary for the physical coupling of two omnidirectional wheeled robots.
Thereby, a constraint optimization problem is formulated for a central \acrshort{mpc}, that also captures the approaching strategy.
The simulation based analysis demonstrates the feasibility of the proposed approach and shows the effectiveness compared to a no coupling approach in a logistics scenario.

Overall, we make three claims:
First, the proposed approach enables in-motion physical coupling of two robots with omni-wheel system dynamics.
Second, our approach is versatile with respect to different initial states. 
Third, we claim that the so-enabled use case in-motion transfer is time and energy saving compared to a non-coupling scenario.

\section{RELATED WORK} \label{sec:Related_Work}
We define logical coupling as the ability of multiple robots to achieve a predetermined spatial configuration, which is characterized by a non-zero inter-robot distance.
This ability can be enabled by solving a formation control problem \cite{ahnFormationControlApproaches2020}.
We define physical coupling as the ability of robots to execute docking operations that, on the one hand, reduce the inter-robot distance up to zero and, on the other hand, physically connect at least two robots.
This ability can be enabled by solving a so-called rendezvous and docking problem \cite{fehseAutomatedRendezvousDocking2003}.
In the context of formation control, a rendezvous problem is the name for a consensus problem, a special class of formation control problems \cite{ohSurveyMultiagentFormation2015}.

\subsection{Logical Coupling of Mobile Ground Robots}
In \cite{ebelOptimizationDrivenControlOrganization2019} an optimal formation for a group of robots is calculated that is used by a \acrfull{dimpc} scheme to enable cooperative transportation of arbitrary shaped objects.
The \acrshort{dimpc} is responsible for maintaining the formation's shape and to move the shape along the desired path.
A leader-follower architecture is used where the leader is a virtual object, and all robots are following this leader with a prescribed pose.
A similar virtual leader approach is used in \cite{egerstedtControlMobilePlatforms2001}, although only one vehicle is controlled.
\cite{rosenfelderCooperativeDistributedNonlinear2022} implements a \acrshort{dimpc} approach with a virtual leader-follower approach that logically couples multiple differentially driven robots.
\cite{stombergCooperativeDistributedMPC2023} implements different \acrshort{dimpc} schemes that couple each robot to its neighbor, following each other with a prescribed offset.
\cite{ebelDistributedModelPredictive2017} implements a \acrshort{dimpc} for multiple robots transporting a plate.
The cost function is designed so that deviations of the desired formation can occur to improve the trajectory tracking error.

These methods ensure formation stability and trajectory or setpoint tracking by controlling a predefined spatial configuration.
However, they do not model any docking constraints or approaching strategies, although they are necessary for successful docking operations \cite{fehseAutomatedRendezvousDocking2003}.
Therefore, these formation control approaches are unsuitable for solving our rendezvous problem to enable in-motion physical coupling.

\subsection{Physical Coupling of Mobile Ground Robots}
Modular autonomous logistic vehicles designed to logically and physically couple are introduced in \cite{hompelTechnicalReportLoadRunner2020}.
A formation controller is implemented that logically couples the robots to enable collaborative transportation.
There are two types of modular vehicles introduced, an active robot that is equipped with a driving unit and a passive carrier that does not have active components.
The active robot is capable of physically coupling with the passive carrier through an electromagnet.
However, they neither investigate approaches for in-motion physical coupling nor for static physical coupling, where the active robot is driving and the other is not.
A modular vehicle for a logistic task with omnidirectional drive and electromagnets to enable static physical coupling is designed in \cite{zhouDesignFabricationControl2022}.
For shape reconfiguration, the coupling point is static and is calculated by the position of the target vehicle to be coupled with.
During reconfiguration, the target vehicle is static while the chasing vehicle is driving.
In \cite{zhouReconfigurableModularVehicle2022}, the trajectory planning is extended with an artificial potential field method that also considers static physical coupling.
Indeed, this approach considers docking constraints, but not for a dynamic setting, where a synchronization of velocities and docking directions should be implemented to prevent frontal collisions.
A leader-follower consensus algorithm with trajectory tracking is implemented in \cite{patilMultiRobotTrajectoryTracking2022}.
The objective is to find a convergence point where a group of robots can rendezvous.
This approach is not applicable as they do not model the necessary docking interfaces.
A PID-based controller is introduced in \cite{yiConfigurationControlPhysical2022} to physically couple two robots in motion.
The drawback of this approach is that there are infeasible initial states if the connection points are on the wrong side of the robot.
The controller cannot plan a trajectory around the other robot to couple from the other side.

These methods also lack the modeling of any docking constraints or approaching strategies, making them unsuitable for enabling in-motion physical coupling.

\subsection{Physical Coupling of Spacecrafts}
Primarily, there is a need for spacecrafts to have the ability to physically couple in motion.
Thus, the rendezvous and docking problem is mainly examined and defined there, even if there are investigations in other domains \cite{jiaRobustDistributedCooperative2024,khanApplicationModularVehicle2023,wuModularAdaptiveAutonomous2021}.
It is a key technology to space missions such as repairing satellites or assembling a space station out of modules \cite{luoSurveyOrbitalDynamics2014}.
The rendezvous and docking is defined in \cite{luoSurveyOrbitalDynamics2014} as "two spacecrafts meet in space with the same velocity and the join into a complex".
Commonly in such space missions, spacecrafts start out of sight of each other, go through several phases to get in close-range, and end in a docking maneuver where they physically couple. 
In such missions, two actors are defined, the chaser, which is the active part and performs the maneuvers, and the target on which the chaser aims to dock \cite{luoSurveyOrbitalDynamics2014,wertzAutonomousRendezvousDocking2003}.
To solve rendezvous and docking problems in space, commonly optimal control methods \cite{zhangTrajectoryOptimizationSpacecraft2022} are used, including \acrshort{mpc} \cite{dicairanoModelPredictiveControl2012,liModelPredictiveControl2017,weissModelPredictiveControl2015,ravikumarAutonomousTerminalManeuver2016}.
\cite{liModelPredictiveControl2017} refers to various papers that propose a \acrshort{mpc} for close-range rendezvous problems because it can handle constraints in a multivariable control system.
For further reading on control methods for space rendezvous and docking is referred to \cite{luoSurveyOrbitalDynamics2014,fehseAutomatedRendezvousDocking2003}.

The approaches for physical coupling of spacecrafts do model docking constraints and approaching strategies, making them suitable as a starting point for our work.
However, these approaches require adaptation to be applied to mobile ground robots, as the boundary conditions differ.
First, these approaches consider the system dynamics of two spacecraft, which are fundamentally different from those of a mobile ground robot.
Secondly, it is necessary to assess which constraints and approaching strategies from space missions are applicable to mobile ground robots.
Thirdly, these approaches assume that only one spacecraft is controllable while the other is passive.
In our case, both mobile ground robots are controllable, which also influences the design of the approaching architecture.  
Given the constraints imposed by the docking interface and the multivariable nature of the control system, employing \acrshort{mpc} approach seems appropriate. 

\section{Constraint Modeling for Omnidirectional-Wheeled robots}\label{sec:Problem_Formulation}
The problem addressed in this paper is to enable in-motion physical coupling for two omnidirectional wheeled robots with an electromagnet as a docking interface.
To enable this ability, we first model each robot's system dynamics to predict the robot's states and plan trajectories.
Afterwards, we adapt the approaching strategy from space rendezvous and docking for mobile ground robots to achieve reliable and unambiguous approaching.
Lastly, all necessary constraints imposed by the docking interface are derived to ensure feasibility for all initial poses.

\subsection{System Dynamics}
Two wheeled mobile robots with the same system dynamics on a two-dimensional free plane are considered.
We assume omnidirectional wheeled robots since they are widespread in logistics \cite{hompelTechnicalReportLoadRunner2020,zhouDesignFabricationControl2022,ebelDistributedModelPredictive2017,eschmannHighAccuracyDataBased2022,colombKonzeptZurIntuitiven2020}.
Each robot $i$ is modeled as a disk with radius $r$.
The pose of robot $i$ is described by the center point of the disk with x-position $p^i_{x} \in \mathbb{R} $, y-position $p^i_{y} \in \mathbb{R}$ and heading $\theta^i_k \in [0,2\pi)$ in global reference system $O_g$ and leads to the state vector
\begin{equation}
    \mathbf{x}^{i}= 
        \begin{bmatrix}
            p^{i}_{x}&
            p^{i}_{y}&
            \theta^{i}
        \end{bmatrix}^{T}
    \label{eq:single_robot_states}
\end{equation}
at time $k$.
Both robots can be controlled individually with translational velocity in the x-direction $v^{i}_{x}$ and the y-direction $v^{i}_{y}$ as well as the angular velocity $\omega^{i}$.
This results in the input vector
    \begin{equation}
    	\mathbf{u}^{i} = \begin{bmatrix}  		
    			v^{i}_{x}&
                v^{i}_{y}&
    			\omega^{i}
    	\end{bmatrix}^{T}
        \label{eq:single_robot_input}
    \end{equation}
of robot $i$ at time $k$.
In real-world scenarios, all actuators are limited due to physical constraints or energy limitations.
Thus, we define the maximal control inputs as
\begin{equation}
    \mathbf{u}^{i} \leq \mathbf{u}^{i}_\text{max}
\end{equation}
The discrete system dynamic for an omnidirectional wheeled mobile robot is derived by Euler-Forward
\begin{equation}
        \begin{aligned}
            \mathbf{x}^{i}_{k+1} = \mathbf{x}^{i}_{k} + \Delta T  \cdot \mathbf{u}^{i}_{k}
        \end{aligned}
        \label{eq:single_robot_system_dynamics}
\end{equation}

\subsection{Docking Interface}
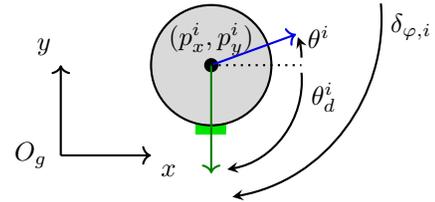
\begin{figure}[h]
\centering
\vspace{-1em}
\begin{tikzpicture}[scale = 4, every node/.style={font=\normalsize}]
    \def\cx{0.5}
    \def\cy{0.5}
    \def\radius{0.2}
    \def\arrowlength{0.3}
    \def\deltaPhi{\arrowlength*2}

    \def\rotation{20} 
    \def\rotateangle{20} 

    \fill[green!90!black] (\cx - 0.05, \cy - \radius+0.1) rectangle (\cx + 0.05, \cy - \radius -0.03) node[left]{};
    
    \draw[->, thick] (0,0.2) -- (\arrowlength,0.2) node[below right] {$x$};
    \draw[->, thick] (0,0.2) -- (0,0.2+\arrowlength) node[above left] {$y$};
    \node (Og) at (-0.1,0.2) {$O_g$};

    \filldraw[fill=gray!30, draw=black, thick] (\cx,\cy) circle(\radius);
    \filldraw[fill=black, draw=black, thick] (\cx,\cy) circle(\radius*0.1) node[above]{$(p^i_x,p^i_y)$};

    \draw[->, thick, blue!90!black] (\cx,\cy) --++ ({\arrowlength*cos(\rotation)}, {\arrowlength*sin(\rotation)}) node[right] {};
    \draw[->, thick, green!50!black] (\cx,\cy) --++  (0, -\arrowlength*1.2) node[below right] {};
    \node at (\cx - 0.15, \cy - 0.1) [above left]{};

    \draw[dotted, thick] (\cx,\cy) --++ (\arrowlength,0);

    \draw[->, thick, >=stealth]
    (\cx + \arrowlength, \cy*1.05)
    arc[start angle=0, end angle=20, radius=0.2]
    node[right] {$\theta^i$};

    \draw[->, thick, >=stealth]
    (\cx + \arrowlength, \cy*0.95)
    node[below right] {$\theta^i_d$}
    arc[start angle=5, end angle=-80, radius=\arrowlength];

    \draw[->, thick, >=stealth]
    ($ (\cx,\cy) + ({\deltaPhi*cos(\rotation)}, {\deltaPhi*sin(\rotation)}) $)
    node[below right] {$\delta_{\varphi,i}$}
    arc[start angle=5, end angle=-80, radius=\deltaPhi] ;


\end{tikzpicture}
\vspace{-3em}
\caption{Illustration of the robot modeling. Robot $i$ is a grey disk with radius $r$ (\protect\greycircle). The heading $\theta^i$ is depicted as a blue arrow ({\color{blue!90!black} $\rightarrow$}). The docking interface is located on the margin of the disk and depicted as a green rectangle ({\color{green!90!black} $\blacksquare$}). The heading of the docking axis is depicted as a green arrow ({\color{green!50!black} $\rightarrow$}). }
\label{fig:robot_modeling}
\end{figure}

The implementation of a coupling interface can be, for example, an electromagnet or a mechanical closure \cite{dokuyucuAchievementsFutureDirections2023}.
We consider an electromagnet as a docking interface and model it as a rectangle on the margin of the robot.
The direction of the coupling interface is set by the parameter $\delta_{\varphi,i} \in [-\pi,\pi)$ that denotes the orientation shift of the coupling interface in reference to the heading $\theta^{i}$ (cf. Fig. \ref{fig:robot_modeling}).
The heading $\theta^i_d \in [0,2\pi)$ of the docking interface is defined by
\begin{equation}
    \theta^i_d = \theta^{i} + \delta_{\varphi,i}
\end{equation}
The center point $\mathbf{p}^i_d \in \mathbb{R}^2$ of the docking interface is defined by
\begin{equation}
    \mathbf{p}^i_{d} = \begin{bmatrix}
        p^{i}_{x} + r \cdot \cos{\theta^i_d} \\
        p^{i}_{y} + r \cdot \sin{\theta^i_d}
    \end{bmatrix}
\end{equation}
Fig. \ref{fig:robot_modeling} illustrates the modeled robot with its docking interface.
The robot is depicted as a grey circle with a black margin.
The docking interface is depicted as a green rectangle on the margin.

\subsection{Approaching Strategy} \label{subsec:Main_Phases}
The approaching strategy is mainly derived from a common rendezvous and docking mission in space, where the approaching strategy can be divided into several phases \cite{luoSurveyOrbitalDynamics2014, fehseAutomatedRendezvousDocking2003, wertzAutonomousRendezvousDocking2003}.
Not all phases are relevant for the considered problem here, as we assume, for example, that the mobile robots are initialized in line of sight, in contrast to a space mission.
Thus, our problem does not cover the phases \textit{launch} and \textit{phasing}, where the main goal is to get into a stable state and in line of sight.
We derive three main phases for the physical coupling of \acrshort{amr}.

The first phase is called \textit{far range rendezvous}, where the goal is to approach the first aim point.
In our case, this is the approaching of the docking axis.
The proposed strategy is that robot 2 aims for the docking axis of robot 1 and concurrently aligns its docking axis to that of robot 1.
This also aligns the coupling interfaces.

The second phase is called \textit{close range rendezvous}, which is divided into the two sub phases \textit{closing} and \textit{final approach}.
The closing reduces the distance between the coupling interfaces and acquires conditions to allow the final approach.
In our context, this phase is about approaching the docking corridor and reducing the relative velocity between the robots.
The final approach is about achieving the docking, where the coupling interfaces touch each other, and the approaching corridor constrains the movements.

The third and last phase is \textit{docking}, where the capture is achieved.
In this context, we assume an electromagnet to achieve docking.

\subsection{Physical Coupling Constraints}
\label{sec:constraints}
We derive the relevant physical coupling constraints mainly from the rendezvous and docking mission from space and adapt them to mobile ground robots.
Concerning the approaching strategy, each phase imposes different constraints that must be modeled and considered in the approach to achieve docking in motion.
Regarding the far range rendezvous phase, we need a description of the docking axis and the relative position of one robot to the docking axis.
For the closing phase, we need a description of the alignment of the docking interface, the relative distance between the robots, and the relative velocity to enable soft-docking.
For the final approach phase, the approaching corridor has to be modeled.
The docking phase itself does not impose any constraints, but prohibits the coupled system to have slack in the coupling constraints afterwards.

\subsubsection{Docking Axis Constraint}
The approaching direction and the position to be approached depend on both robots' positions and the orientation of the installed coupling interface.
The docking axis constraint is introduced to ensure that the robots couple from a feasible direction (cf. Fig. \ref{fig:robot_corridor}).
First, we define the angle $\varphi_{12} \in [0,2\pi)$ in reference to the x-axis of the vector that starts at robot 1 and ends at robot 2. 
\begin{equation}
    \varphi_{12} = \mathrm{atan2}(p^2_{y}-p^1_{y},p^2_{x}-p^1_{x})
    \label{eq:phi_robot_1_to_2}
\end{equation}
Thereby, it is possible to describe the deviation of the position of robot 2 from the docking axis of robot 1 by building the difference between $\theta^{1}_{d}$ and $\varphi_{12}$.
This deviation has to be zero for physical coupling and leads to the constraint
\begin{equation}
\label{eq:constraint_approaching_angle}
 \underbrace{\theta^{1}_{d}-\varphi_{12}}_{\Rom{1}} = 0
\end{equation}

\subsubsection{Alignment Constraint}
While the docking axis constraint \eqref{eq:constraint_approaching_angle} only aligns the position of robot 2 with the docking axis of robot 1, the alignment of the docking interfaces is lacking.
Following \cite{yiConfigurationControlPhysical2022}, a heading alignment constraint can be used to control the alignment of the coupling interfaces.
As an alignment of the coupling interface is obtained, the heading of the docking interface is used instead of the heading of the robots.
The coupling interfaces are aligned if their docking axes point at each other.
For the mathematical modeling, the difference between the docking axis heading $\Delta \theta_d$ is mapped to values from $[0,\pi]$.
\begin{equation}
     \Delta \theta_d = \theta^1_d  - \theta^2_d
\end{equation}
The docking axes are pointing at each other, if $\Delta \theta_d$ equals $\pi$, which leads to the alignment constraint 
\begin{equation}
    \underbrace{\Delta \theta_d - \pi}_{\Rom{2}} = 0
\end{equation}

\subsubsection{Relative Distance Constraint}
The relative distance between the robots is defined by the distance of their center point.
To enable the docking operation, the relative distance between both robots has to be the sum of the distance of the modeled disk radius $r$ and thus leads to
\begin{equation}
\label{eq:constraint_relative_distance}
     \underbrace{(p^1_{x}-p^2_{x})^2+(p^1_{y}-p^2_{y})^2 - \delta_r}_{\Rom{3}} = 0
\end{equation}
\cite{yiConfigurationControlPhysical2022} proposes a similar constraint for mobile ground robots.
The parameter $\delta_r \in [0,\infty)$ defines the relative distance.

\subsubsection{Soft-Docking Constraint}
In order to prevent damage, the force transmission should be as small as possible which is why the relative velocity and acceleration of the robots have to be constrained to a minimum while docking.
Following \cite{dicairanoModelPredictiveControl2012}, a soft-docking constraint is formulated to lower the relative velocity of the robots as they are approaching each other with
\begin{equation} \label{eq:constraint_soft_docking}
    \underbrace{(v^1_{x}-v^2_{x})^2+(v^1_{y}-v^2_{y})^2}_{\Rom{4}} = 0
\end{equation}

\subsubsection{Approaching Corridor}
\begin{figure}[h]
\centering
\vspace{-2em}
\begin{tikzpicture}[scale = 4, every node/.style={font=\normalsize}]
    \def\cx{0}
    \def\cy{0.5}
    \def\cxT{1}
    \def\cyT{0.8}
    
    \def\radius{0.2}
    \def\arrowlength{0.3}
    \def\deltaPhi{\arrowlength*2}

    \def\rotation{-45} 
    \def\rotaterectangle{45} 
    \def\rotateangle{-40} 

    \def\gap{30} %
    \def\rCorridor{2}

    
    
    \begin{scope}[rotate around={\rotaterectangle:(\cx,\cy)}]
      \fill[green!90!black] 
        (\cx - 0.05, \cy - \radius + 0.1) 
        rectangle 
        (\cx + 0.05, \cy - \radius - 0.03);
    \end{scope}

    \filldraw[fill=gray!30, draw=black, thick] (\cx,\cy) circle(\radius);
    \filldraw[fill=black, draw=black, thick] (\cx,\cy) circle(\radius*0.1) node[above]{};

    \begin{scope}[rotate around={\rotaterectangle:(\cxT,\cyT)}]
      \fill[green!90!black] 
        (\cxT - 0.05, \cyT - \radius + 0.1) 
        rectangle 
        (\cxT + 0.05, \cyT - \radius - 0.03);
    \end{scope}

    \filldraw[fill=gray!30, draw=black, thick] (\cxT,\cyT) circle(\radius);
    \filldraw[fill=black, draw=black, thick] (\cxT,\cyT) circle(\radius*0.1) node[above]{};

    \draw[->, thick, green!50!black] (\cx,\cy) --++  ({\arrowlength*cos(\rotation)}, {\arrowlength*sin(\rotation)}) node[below right] {};
    \draw[->, thick, green!50!black] (\cxT,\cyT) --++  ({\arrowlength*cos(\rotation)}, {\arrowlength*sin(\rotation)}) node[below right] {};

\begin{scope}
  \clip (\cx,\cy)
    -- ++({\rotation - \gap}:\radius*\rCorridor)
    arc[start angle={\rotation - \gap}, end angle={\rotation + \gap}, radius=\radius*\rCorridor]
    -- cycle;

  \fill[green!60!black, opacity=0.3] (\cx,\cy) circle(\radius*\rCorridor);
\end{scope}

\begin{scope}
  \clip (\cx,\cy)
    -- ++({0}:\radius*\rCorridor)
    arc[start angle={0}, end angle={\rotation + \gap}, radius=\radius*\rCorridor]
    -- cycle;

  \fill[red!60!black, opacity=0.3] (\cx,\cy) circle(\radius*\rCorridor);
\end{scope}

\begin{scope}
  \clip (\cx,\cy)
    -- ++({\rotation - \gap}:\radius*\rCorridor)
    arc[start angle={\rotation - \gap}, end angle={-360}, radius=\radius*\rCorridor]
    -- cycle;

  \fill[red!60!black, opacity=0.3] (\cx,\cy) circle(\radius*\rCorridor);
\end{scope}



    \draw[dotted, thick] (\cx,\cy) --++ (\arrowlength,0);
    \draw[dotted, thick] (\cxT,\cyT) --++ (\arrowlength,0);
    
    \draw[->, thick, >=stealth]
    (\cx + \arrowlength, \cy*0.95)
    node[below right] {$\theta^1_d$}
    arc[start angle=0, end angle=\rotateangle, radius=\arrowlength];

    \draw[->, thick, >=stealth]
    (\cxT + \arrowlength, \cyT*0.95)
    node[below right] {$\theta^2_d$}
    arc[start angle=0, end angle=\rotateangle, radius=\arrowlength];

    \draw[dotted, thick] (\cx,\cy) -- (\cxT,\cyT) node[midway, above] {};

    \draw[->, thick, >=stealth]
    ($ (\cx,\cy) + ({\deltaPhi*cos(15)}, {\deltaPhi*sin(15)}) $)
    node[below right] {$\Delta \theta_{\varphi}$}
    arc[start angle=5, end angle=-45, radius=\deltaPhi] ;

    \draw[->, thick, >=stealth]
    (\cx + \arrowlength, \cy*1.05)
    node[right] {$\varphi_{12}$}
    arc[start angle=5, end angle=15, radius=\arrowlength];


\draw[->, thick, >=stealth]
  (\cx,\cy) ++(-47:\arrowlength*1.05)  
  arc[start angle=-47, end angle=-75, radius=\arrowlength]
  node[pos=0.5, below right] {$\Delta_{\varphi}$};

\end{tikzpicture}
\vspace{-3em}
\caption{Illustration of the approaching corridor and the docking axis displacement. The robots are depicted as grey disk (\protect\greycircle). The docking interface is located on the margin of the disk and depicted as a green rectangle ({\color{green!90!black}$\blacksquare$}). The heading of the docking axis is depicted as a green arrow ({\color{green!50!black} $\rightarrow$}). The approaching corridor is depicted as a green zone around around the docking axis(\protect\greenzone). The collision avoidance zone is depicted as a red zone (\protect\redzone) }
\label{fig:robot_corridor}
\end{figure}
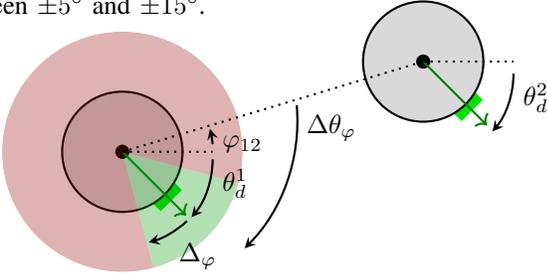

To prevent robot 2 from colliding with robot 1 during approaching the docking axis of robot 1, collision avoidance is considered in the constraints.
Therefore, a disk with radius $r_{ca}$ is defined around the center point of robot 1 that is prohibited to enter (cf. Fig. \ref{fig:robot_corridor}).
This means that the euclidean distance between the center points of both robots has to be greater than $r_{ca}$.
This is expressed via the collision avoidance term $\alpha_{ca}$ with
\begin{equation}
     \alpha_{ca} = ((p^1_{x}-p^2_{x})^2+(p^1_{y}-p^2_{y})^2-r_{ca}^2) \geq 0
     \label{eq:collision_avoidance}
\end{equation}
Having only this constraint would be too restrictive, because a physical coupling would be impossible.
In the space domain, existing approaches create an interruption in the collision avoidance perimeter at the designated docking axis where it is allowed to enter \cite{zhangTrajectoryOptimizationSpacecraft2022}.
The aperture angle of the corridor is defined around the docking axis by the parameter $\Delta_\varphi$ (cf. Fig. \ref{fig:robot_corridor}).
The deviation from the docking axis is defined by $\Delta \theta_\varphi$ and mapped to [0,$\pi$].
\begin{equation}
    \Delta \theta_\varphi = \theta^1_d  - \varphi_{12}
\end{equation}
For reasons of numerical stability, a continuously differentiable function is preferred over a discrete function in the constraints.
Thus, tangens hyperbolicus is used instead of a step function to decide whether robot 2 is inside or outside of the corridor.
For all deviations from the docking axis that are smaller than $\Delta_\varphi$, the term within $\tanh$ becomes negative and thus tanh becomes -1.
For all deviations from the docking axis that are greater than $\Delta_\varphi$, the term within $\tanh$ becomes positive and thus tanh becomes 1.
The hyperparameter $l$ is used to tune the gradient of the tangens hyperpolicus.
While increasing $l$, the tangens hyperbolicus approximates a step function.
\begin{equation}
     \alpha_{ce} = (1+\tanh(l\cdot(\Delta \theta_\varphi-\Delta_\varphi)))
\label{eq:collision_embrace_cone}
\end{equation}
Thereby, the docking corridor term $\alpha_{dc}$ is zero if robot 2 is within the cone and two if robot 2 is outside of the cone.
In order to connect \eqref{eq:collision_avoidance} with \eqref{eq:collision_embrace_cone} we have to multiply $\alpha_{ca}$ with $\alpha_{ce}$.
\begin{equation}
     0 \leq \alpha_{ca}\cdot\frac{1}{2}\alpha_{ce}
\end{equation}
As soon as robot 2 is inside the cone $\alpha_{ce}$ becomes zero, and so $\alpha_{ca}$ is inactive.
While robot 2 is outside the cone $\alpha_{ce}$ is 2 and $\alpha_{ca}$ is active.

According to \cite{fehseAutomatedRendezvousDocking2003}, the half cone angle of the approach corridors are arbitrary but for spacecrafts the angle may vary between $\pm 5 ^{\circ}$ and $\pm 15 ^{\circ}$.

\section{Model Predictive Controller Design} \label{sec:Approach}
Our approach involves designing a central model predictive controller that calculates optimal trajectories based on physical coupling.
We now define the cost function and constraints so that the physical coupling constraints are met, the approaching strategy is implemented, and the user-given waypoints are approached physically coupled.

\subsection{Slack Variables}
Considering the approaching strategy \ref{subsec:Main_Phases}, different constraints should be met in different phases. 
In the final approach phase, right before docking, all constraints from all phases have to be met.
However, all constraints cannot be fulfilled from the very beginning, thus slack variables $\boldsymbol{\epsilon} \in \mathbb{R}^4$ are introduced.
 \begin{equation}
        \boldsymbol{\epsilon}_k = \begin{bmatrix}
        \epsilon_{\delta_r}&
        \epsilon_{\delta_\theta}&
        \epsilon_{\delta_v}&
        \epsilon_{\delta_\varphi}
        \end{bmatrix}^{T}
    \end{equation}
Adding slack variables in the constraints can soften state constraints so that the slack variable behaves like a decision variable \cite{rawlingsMPCbook2ndedition4thprinting2022}.
The individual constraints can be prioritized depending on how the slack variables are penalized in the cost function.
The physical coupling constraint equations \eqref{eq:constraint_approaching_angle} to \eqref{eq:constraint_soft_docking} are augmented by the inclusion of their related slack variable.
If $\epsilon$ takes the value zero, the constraint is satisfied.
The greater the value of $|\epsilon|$, the more the constraint is violated.
The slack variable gives the \acrshort{mpc} a degree of freedom to drive the system into a state where the constraints are satisfied by minimizing epsilon in the cost function.
To ensure that already satisfied constraints are maintained, we have to constrain the slack variables to a maximum value $\boldsymbol{\epsilon}_{max} \in \mathbb{R}^4$
\begin{equation}
    \left|\boldsymbol{\epsilon}\right| \leq \boldsymbol{\epsilon}_\text{max}.
\end{equation}

\subsection{Modeling Central System Dynamics}
We consider a central \acrshort{mpc} architecture with global information.
For the dynamical model, all states of each robot are considered.
Therefore, we define the extended state vector $\mathbf{z}_{k} \in \mathbb{R}^{6}$ and the extended input vector $\boldsymbol{\nu}_{k} \in \mathbb{R}^{6}$ with the central discrete system dynamics $\mathbf{z}_{k+1} \in \mathbb{R}^{6}$.

\[
\begin{minipage}{0.45\linewidth}
    \begin{equation}
    	\mathbf{z}_{k} = \left(
    		\begin{aligned}
    			\mathbf{x}^1_{k}\\
                \mathbf{x}^2_{k}
    		\end{aligned}
    	\right)
        \label{eq:multi_robot_sates}
    \end{equation}
\end{minipage}
\hfill
\begin{minipage}{0.45\linewidth}
    \begin{equation}
    	\boldsymbol{\nu}_k = \left(
    		\begin{aligned}
    			\mathbf{u}^1_{k}\\
                \mathbf{u}^2_{k}
    		\end{aligned}
    	\right)
    \end{equation}
\end{minipage}
\]

    \begin{equation}
        \mathbf{z}_{k+1} = \mathbf{f}(\mathbf{z}_{k},\boldsymbol{\nu}_{k}) = \mathbf{z}_{k} + \Delta T  \cdot \boldsymbol{\nu}_k
    \end{equation}

\subsection{Cost Function}
The objective of the \acrshort{mpc} is to minimize the slack variables that depict the error of the coupling constraints and to minimize the distance to a common way point.
So that the robots arrive at the waypoint physically coupled without stopping in between, we have to penalize the error and thus the slack variables of the coupling constraints more than the distance to the waypoint.
Consequently, we introduce the parameter vector $\boldsymbol{\lambda} \in \mathbb{R}^6$ with which we can adjust the weighting of the cost function.
\begin{equation}
    \boldsymbol{\lambda} = \begin{bmatrix}
         \lambda_{\delta_r}&
         \lambda_{\delta_\theta}&
         \lambda_{\delta_v}&
         \lambda_{\delta_\varphi}&
         \lambda_j&
         \lambda_\omega
    \end{bmatrix}^{T}
\end{equation}
To satisfy every constraint, all slack variables have to be zero.
Hence, the cost function for the coupling constraint $L_{cc}$ is defined as the parameterized weighted sum of the squared slack variable over all timesteps N within the time horizon.
\begin{equation}
    \begin{aligned}
        L_{cc}(\boldsymbol{\epsilon}_k) =  &\lambda_{\delta_r}\cdot\sum_{k=1}^N \epsilon_{\delta_r\text{, k}}^2
            &+   \lambda_{\delta_\Theta}\cdot\sum_{k=1}^N \epsilon_{\delta_\Theta\text{, k}}^2 \\
            +  &\lambda_{\delta_v}\cdot\sum_{k=1}^N \epsilon_{\delta_v \text{,k}}^2
            &+   \lambda_{\delta_\varphi}\cdot\sum_{k=1}^N \epsilon_{\delta_\varphi\text{, k}}^2\\
    \end{aligned}
\end{equation}
To smooth the input variables the derivatives of the inputs are penalized via
\begin{equation}
    \begin{aligned}
            L_{in}(\boldsymbol{\nu}_k) =  &\lambda_j\cdot\sum_{k=0}^{N-1} (\ddot{v}^1_{x,k} )^2 +  (\ddot{v}^1_{y,k} )^2 + (\ddot{v}^2_{x,k} )^2 +  (\ddot{v}^2_{y,k} )^2\\
            &\lambda_\omega\cdot\sum_{k=0}^{N-1} (\dot{\omega}^1_{k} )^2 +  (\dot{\omega}^1_{k} )^2
    \end{aligned}
\end{equation}
The distance between robot 1 and the common waypoint is defined by the euclidean distance.
We use the squared distance, because it provides better numerical performance while not loosing its interpretation.
The costs for arriving at the waypoint $L_{g}$ are formulated as end costs with the weighting parameter $\boldsymbol\lambda_{g} \in \mathbb{R}^6$ 
\begin{equation}
    \begin{aligned}
        L_g (\mathbf{z}_{k}) = \boldsymbol\lambda_{g}\cdot(\mathbf{z}_N-\mathbf{z}_{soll})^2 \\
    \end{aligned}
\end{equation}
The total costs $L$ result from the sum of coupling costs and waypoint costs.
\begin{equation}
    L (\mathbf{z}_{k},\boldsymbol{\nu}_{k}, \boldsymbol{\epsilon}_k) = L_{cc}(\boldsymbol{\epsilon}_k) + L_{in}(\boldsymbol{\nu}_k) + L_g (\mathbf{z}_{k})
\end{equation}

\subsection{Optimization Problem}
Putting all together, the \acrshort{mpc} minimizes at each time step the cost function $L$ for a time horizon N as defined in \eqref{eq:mpc}.
The cost function is subject to the central system dynamics $\mathbf{z}_{k+1}$ and the physical coupling constraints.
The optimized variables are the states $\mathbf{z}_{1 \rightarrow N}^* \in \mathbb{R}^{6 \times N }$, the input variables $\boldsymbol{\nu}_{0 \rightarrow N-1}^* \in \mathbb{R}^{6 \times N }$ and the slack variables $\boldsymbol{\epsilon}_{1 \rightarrow N}^* \in \mathbb{R}^{4 \times N }$.
Hence, the optimization problem is as follows.
\begin{equation}
\label{eq:mpc}
\begin{aligned}
		  \underset{\mathbf{z}_{k},\boldsymbol{\nu}_{k}, \boldsymbol{\epsilon}_k}{\arg\min \text{ }} & L (\mathbf{z}_{k},\boldsymbol{\nu}_{k}, \boldsymbol{\epsilon}_k) \\
		\text{subj. to } & \mathbf{z}_{k+1} = \mathbf{f}(\mathbf{z}_{k},\boldsymbol{\nu}_{k}) & k= 0,...,N-1\\
		& \mathbf{z}_{0} \in \mathbb{R}^6 &\\
		& \Rom{1} = \epsilon_{\delta_{\varphi},k} & k= 1,...,N\\		
        & \Rom{2} = \epsilon_{\delta_\theta,k} & k= 1,...,N\\
		& \Rom{3}=\epsilon_{\delta_r,k} & k= 1,...,N\\
        & \Rom{4} = \epsilon_{\delta_v,k} & k= 1,...,N\\
        & \left|\boldsymbol{\epsilon}\right| \leq \boldsymbol{\epsilon}_\text{max}  & k= 1,...,N\\
        & \boldsymbol{\nu}_k \leq \boldsymbol{\nu}_\text{max}  & k= 0,...,N-1\\
        & 0 \leq \alpha_{ca}\cdot\frac{1}{2}\alpha_{ce} & k= 1,...,N\\
\end{aligned}
\end{equation}

\section{EVALUATION}\label{sec:Evaluation}
This section evaluates the proposed \acrshort{mpc} motion planning approach to enable physical in-motion coupling of \acrshort{amr}. 
We first describe the experimental setup used to validate the claims and subsequently present the simulative results. 
In addition, we discuss and interpret the results and findings.

\subsection{Experimental Setup}
For each claim, we consider one experiment. 
In the first experiment, we validate the claim that the two omnidirectional robots can physically couple in motion before reaching the goal. 
Therefore, we set the goal state for robot 1 to $[4,0,0]$.
The goal state of robot 2 results from the coupling constraints to $[4,2,0]$.
Robot 1 starts at $[0,-2,0]$ and Robot 2 at $[0,2,0]$, and they are not physically coupled in the beginning, but the docking axes are aligned.
This represents the simplest case, and the robots should couple before arriving at their goal point.

The second experiment validates the claim that the approach can handle a scenario where the docking axes are not aligned initially.
We use the same goal states as in the first experiment, but the start states of the robots are interchanged.

The third experiment validates the claim that the enabled in-motion transfer saves time and energy for a transportation scenario in a logistic context.
Therefore, we define two scenarios, one without coupling and thus without in-motion transfer and the other with physical coupling and in-motion transfer (cf. Fig. \ref{fig:paper_figure}).
Robot 2 carries two packages, one with destination $A = (8,2)$ and the other with destination $B = (8,-2)$.
Robot 1 only carries a package with destination $B$.
The initial states of the robot are the same as in experiment 1.
For both scenarios, the first waypoint is $[2,0]$ for robot 1 and $[2,1]$ for robot 2, as we assume the robots are driving out a shelf row.
In the approach with coupling, robot 2 transfers the package for destination $B$ to robot 1.
The scenarios end if both robots have delivered their packages.

In all experiments, the docking axis for robot 1 is set to $\delta_{\varphi,1} = 90°$ and for robot 2 to $\delta_{\varphi,2} = -90°$.
Both robots are modeled as a disk with $r = 0.1$ and thus $\delta_r = 0.2$.
The approaching corridor is set to $\Delta_\varphi = 15°$.
Considering the approaching strategy the weighting of the cost function is set to $\boldsymbol{\lambda} = [30,1000,1,200,0.1,1]^T$.
The \acrshort{mpc} hyperparameters are the time horizon $T = 5$ and the steps $N=20$.
The end costs are set to $\boldsymbol{\lambda}_g = [1,1,200,1,1,200]$.
The algorithm is implemented with python and casadi \cite{anderssonCasADiSoftwareFramework2019} and solved with ipopt.

\subsection{Results}
\begin{figure*}[!tb]
\vspace{-2em}
  \centering
  \subfloat[Global Position]{%
    \includegraphics[width=0.2\linewidth]{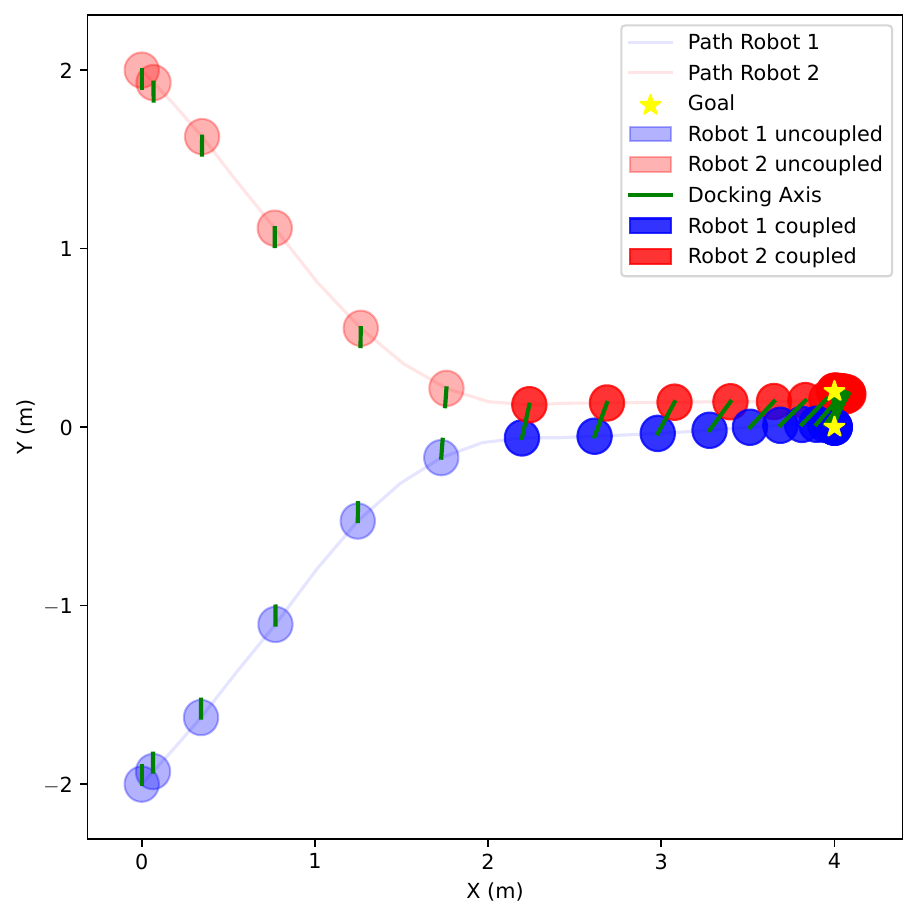}
    \label{fig:results_exp1_glob_pos}%
  }
  \subfloat[Slip Docking Axis]{%
    \includegraphics[width=0.2\linewidth]{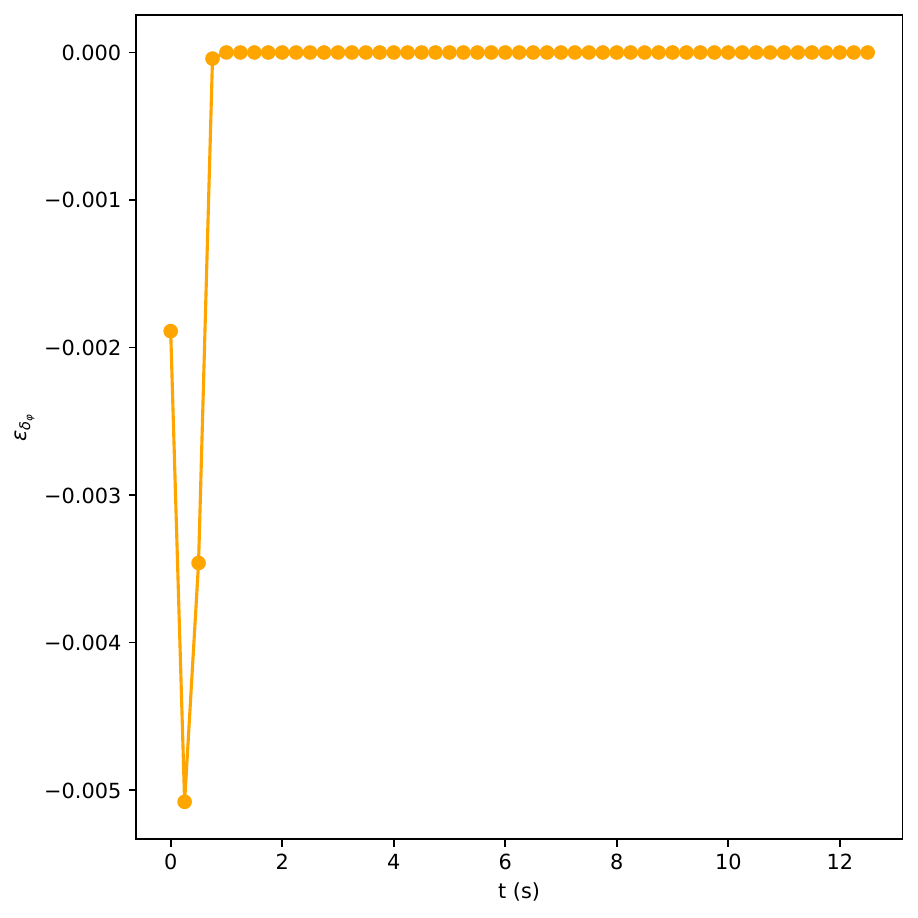}
    \label{fig:results_exp1_slip_docking}%
  }
  \subfloat[Slip Alignment Docking Interface]{%
    \includegraphics[width=0.2\linewidth]{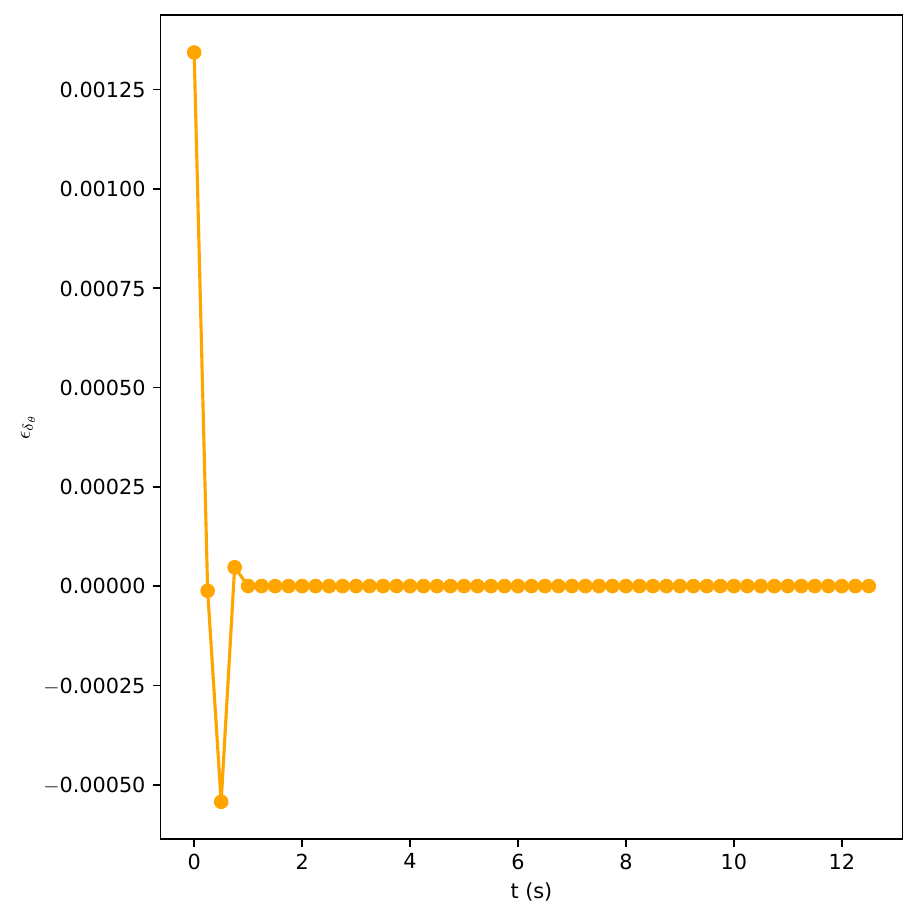}
    \label{fig:results_exp1_slip_alignment}%
  }
    \subfloat[Slip Relative Distance]{%
    \includegraphics[width=0.2\linewidth]{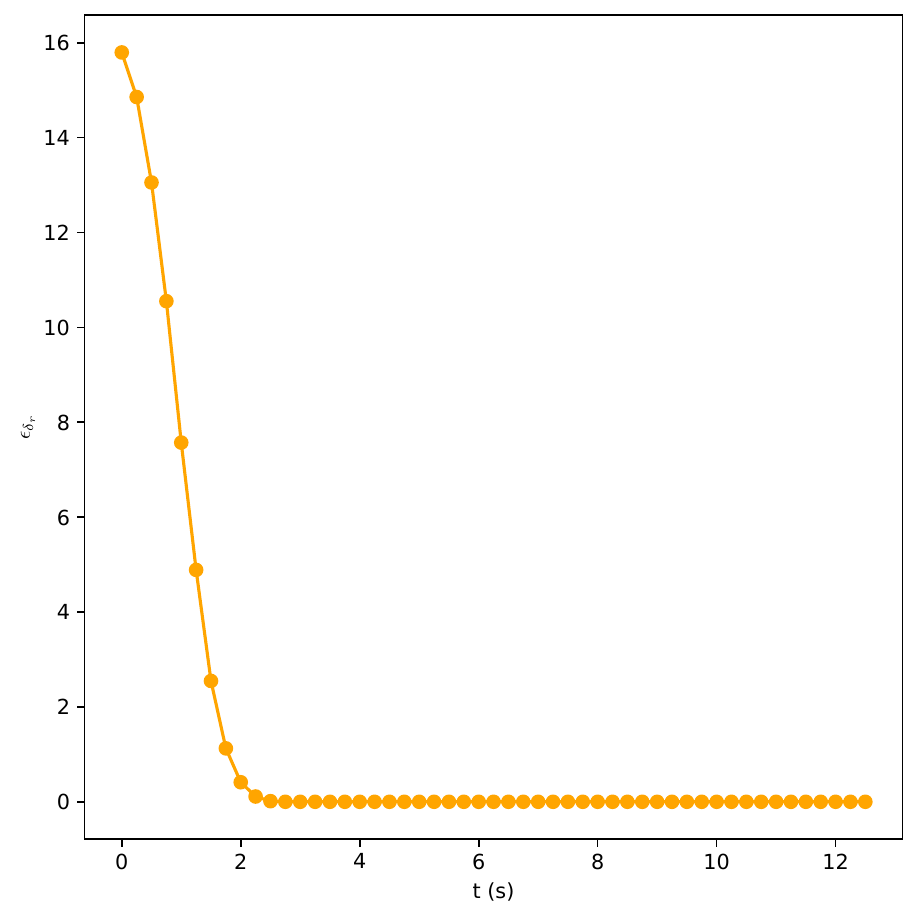}
    \label{fig:results_exp1_slip_distance}%
  }
  \subfloat[Slip Soft-Docking]{%
    \includegraphics[width=0.2\linewidth]{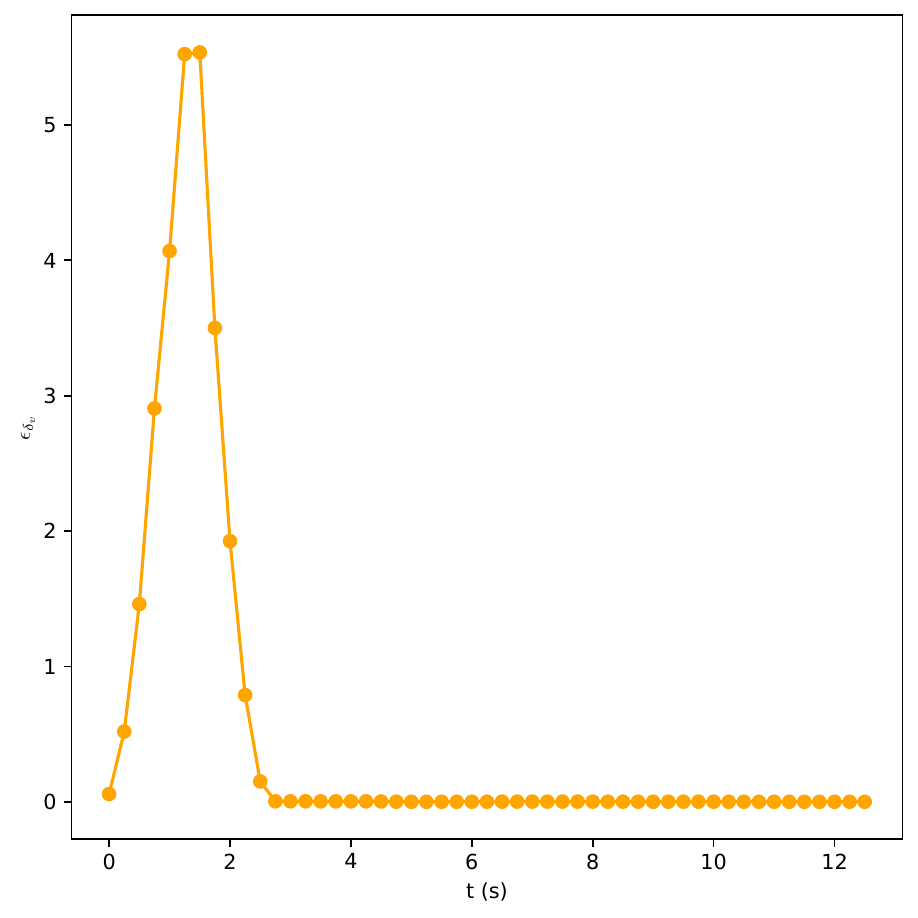}
    \label{fig:results_exp1_slip_soft-docking}%
  }
  \caption{Experiment 1 - The robots' docking axes are perfectly aligned at the beginning. The goal is that the robots satisfy the coupling constraints before arriving at the goal point and thus physically couple in motion.}
  \label{fig:results_exp1}
\end{figure*}

The results of experiment 1 are depicted in Fig. \ref{fig:results_exp1}.
Fig. \ref{fig:results_exp1_glob_pos} shows the temporal course of the global position of robot 1 depicted in blue and robot 2 depicted in red.
The docking axis is marked in green, and the goal position is depicted as a yellow star.
The robots are displayed in non-transparent color once they are physically coupled.
Both robots start at their initial position and drive to their goal point while satisfying the coupling constraints.
This is depicted by the decreased value of the slip variables over time.
The slip in the docking axis (cf. Fig. \ref{fig:results_exp1_slip_docking}) and the alignment (cf. Fig. \ref{fig:results_exp1_slip_alignment}) constraint go towards zero first.
Afterwards, the slip in the relative distance (cf. Fig. \ref{fig:results_exp1_slip_distance}) goes towards zero.
The slip in the soft-docking constraint is increasing first before decreasing to zero (cf. Fig. \ref{fig:results_exp1_slip_soft-docking}).
After 2 seconds, both robots are physically coupled and are driving to their goal state together.

\begin{figure*}[!tb]
\vspace{-2em}
  \centering
  \subfloat[Global Position]{%
    \includegraphics[width=0.2\linewidth]{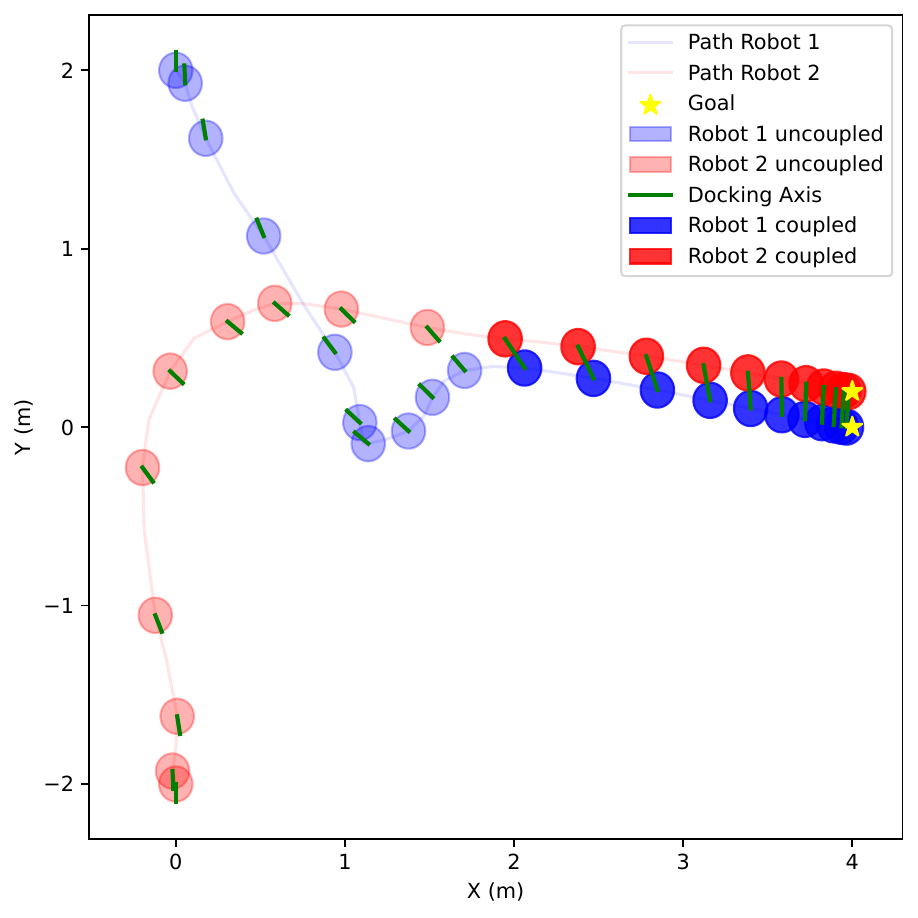}
    \label{fig:results_exp2_glob_pos}%
  }
  \subfloat[Slip Docking Axis]{%
    \includegraphics[width=0.2\linewidth]{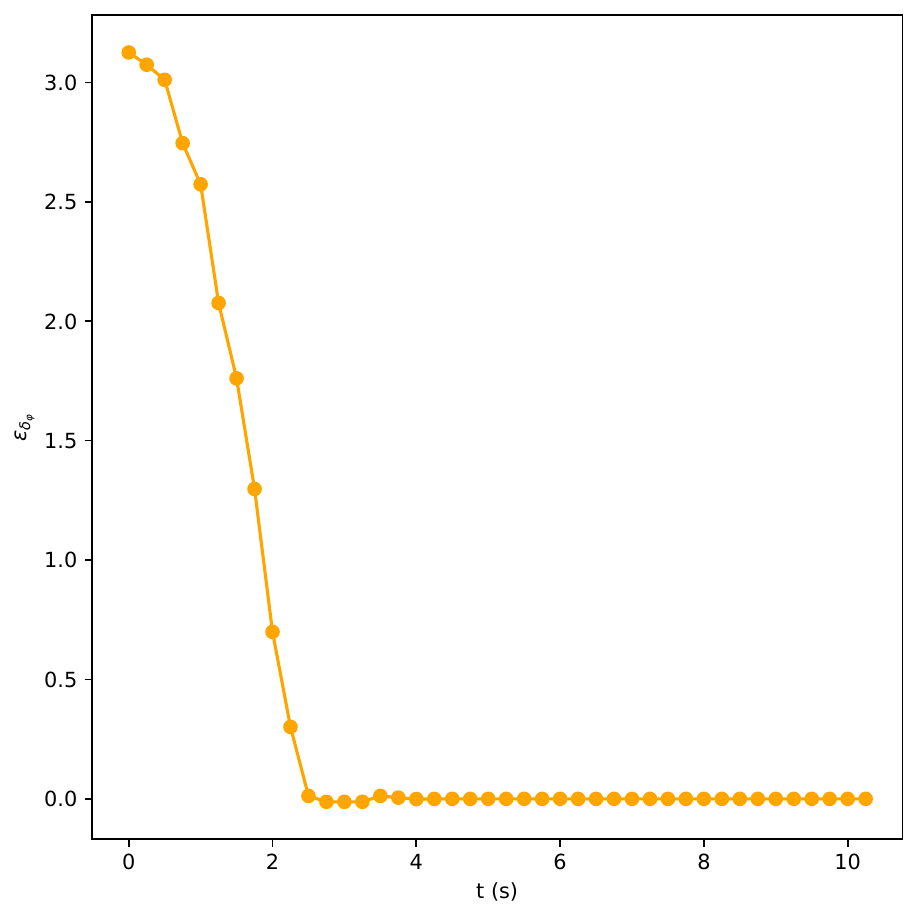}
    \label{fig:results_exp2_slip_docking}%
  }
  \subfloat[Slip Alignment Docking Interface]{%
    \includegraphics[width=0.2\linewidth]{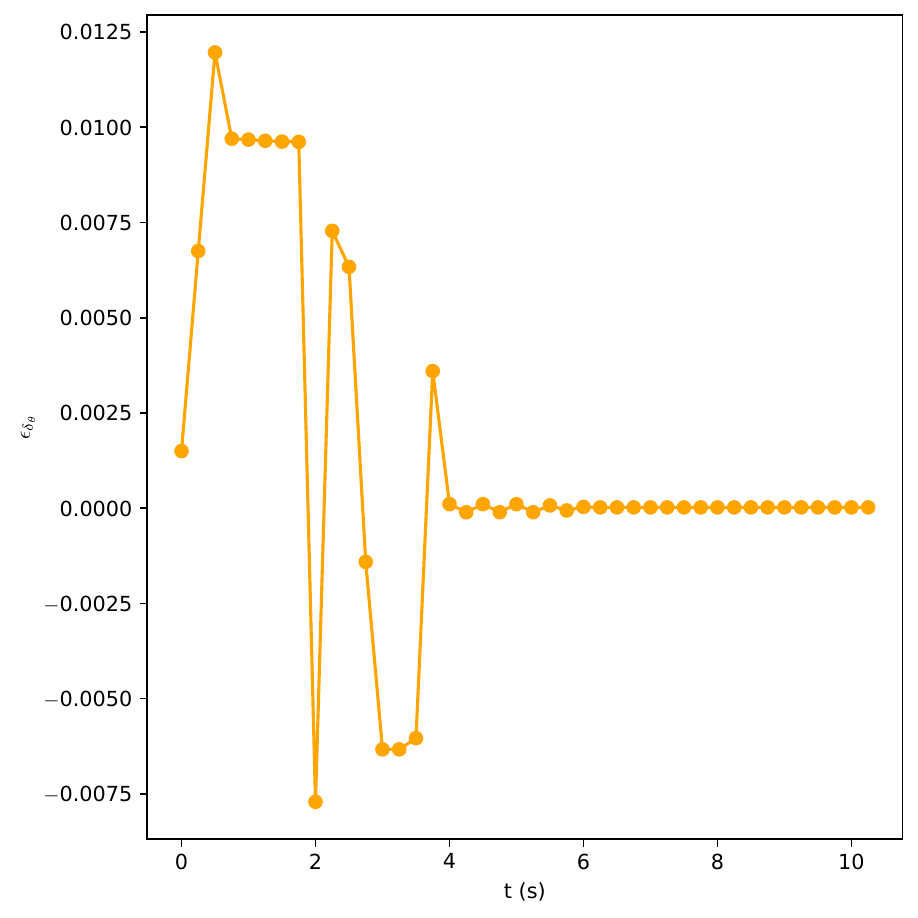}
    \label{fig:results_exp2_slip_alignment}%
  }
    \subfloat[Slip Relative Distance]{%
    \includegraphics[width=0.2\linewidth]{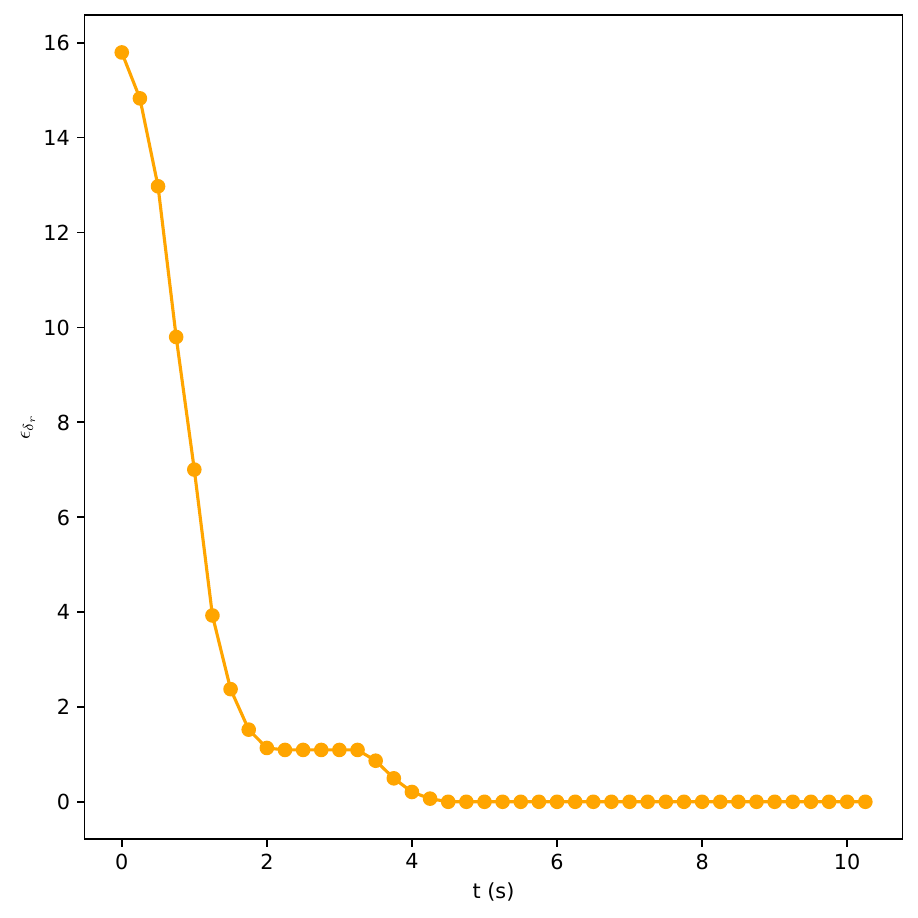}
    \label{fig:results_exp2_slip_distance}%
  }
  \subfloat[Slip Soft-Docking]{%
    \includegraphics[width=0.2\linewidth]{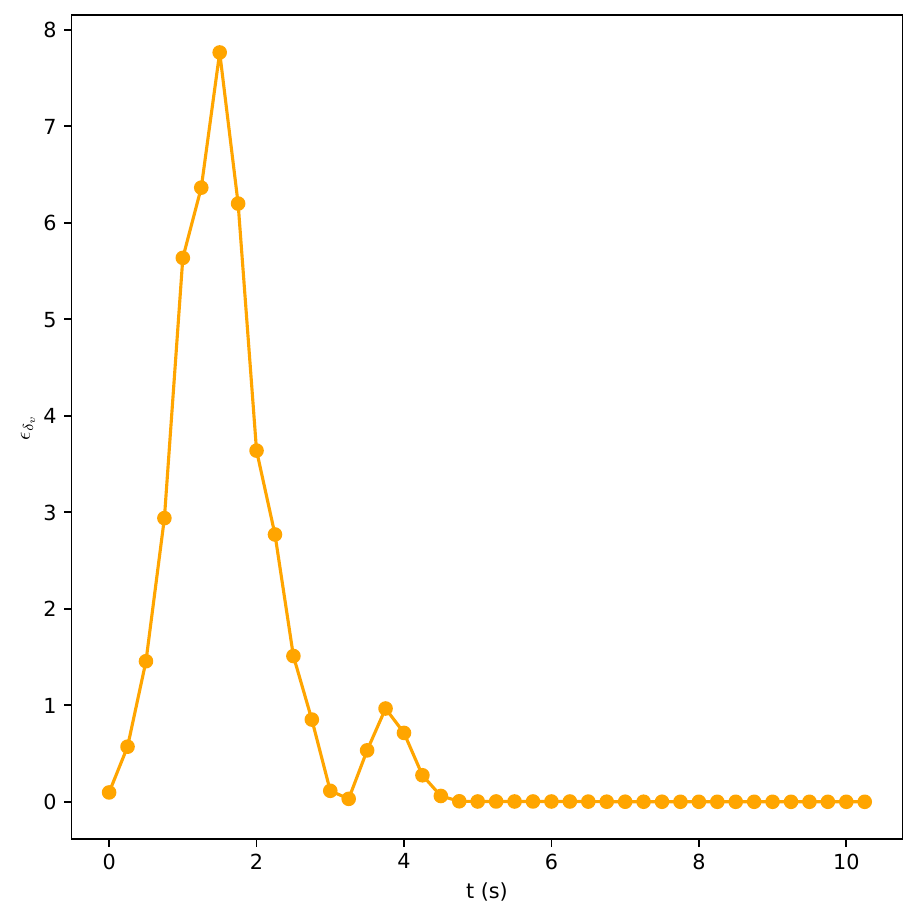}
    \label{fig:results_exp2_slip_soft-docking}%
  }
  \caption{Experiment 2 - The robots' docking axes are not aligned at the beginning. The goal is that the robots satisfy the coupling constraints before arriving at the goal point and thus physically couple in motion.}
  \label{fig:results_exp2}
\end{figure*}

The results of experiment 2 are depicted in Fig. \ref{fig:results_exp2}.
Fig. \ref{fig:results_exp2_glob_pos} shows the temporal course of the global position of the robots analogous to Fig. \ref{fig:results_exp1_glob_pos}.
The slip of the docking axis is monotonously decreasing (cf. Fig.\ref{fig:results_exp2_slip_docking}).
The slip of the alignment of the docking axes is oscillating around zero before converging to it (cf. Fig.\ref{fig:results_exp2_slip_alignment}).
The relative distance is decreasing first, but after 2 seconds it remains on a plateau before it decreases towards zero after 3.5 seconds (cf. Fig.\ref{fig:results_exp2_slip_distance}).
The slip of the soft-docking constraint shows two hills, one after 1.5 seconds and the other after 3.75 seconds (cf. Fig. \ref{fig:results_exp2_slip_soft-docking}).

\begin{figure}[!tb]
\vspace{-2em}
  \centering
  \subfloat[No coupling]{%
    \includegraphics[width=0.5\linewidth]{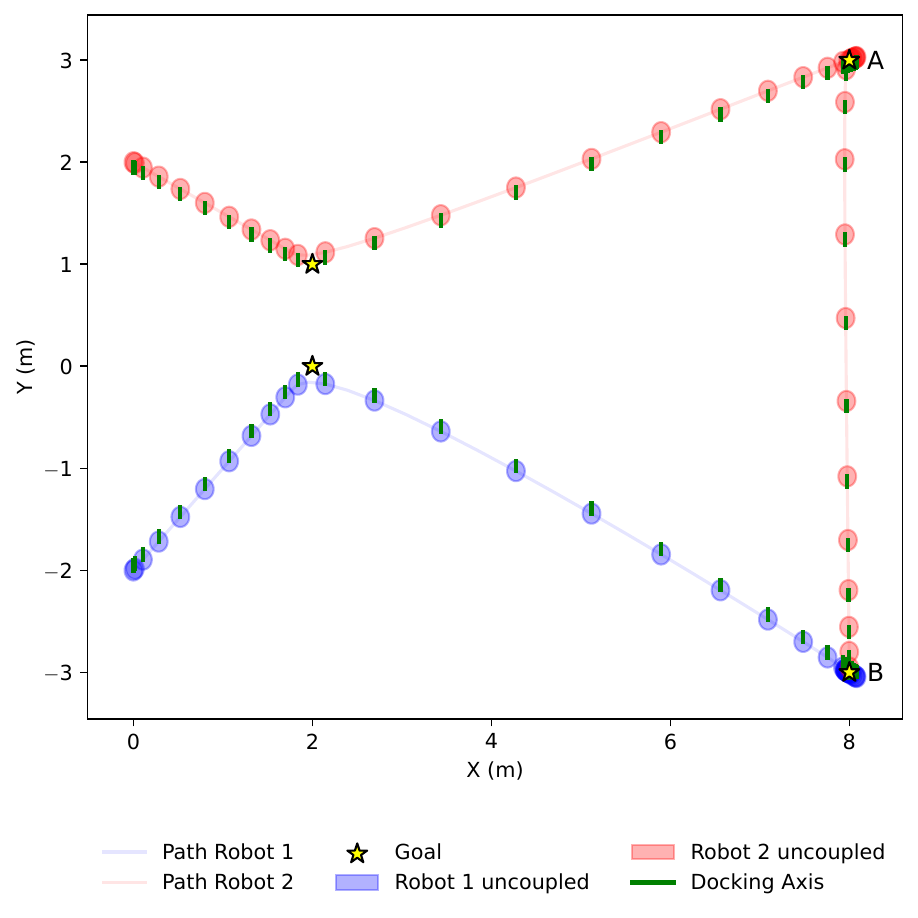}
    \label{fig:results_exp3_no_coupling}%
  }
  \subfloat[Rendezvous and Docking with in-motion transfer]{%
    \includegraphics[width=0.5\linewidth]{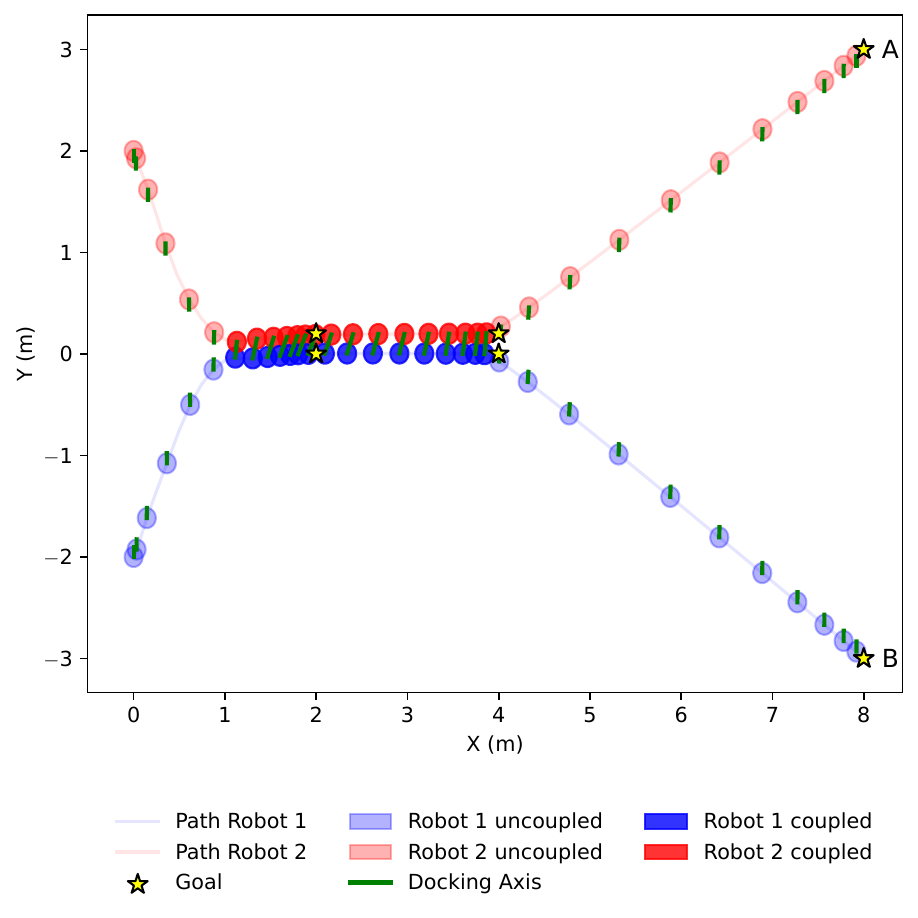}
    \label{fig:results_exp3_with_coupling}%
  }
  \caption{Experiment 3 - A comparison is performed between a no coupling scenario and a physically coupling scenario with in-motion transfer. Robot 1 only has to deliver a package to point $B$. Robot 2 has to deliver two packages, one to point $A$ and $B$. In the no coupling scenario, robot 2 has to drive to both destinations, whereas in the physically coupling scenario, robot 2 only has to drive to $A$ because the package can be transferred to robot 1 due to the so enabled in-motion transfer.}
  \label{fig:results_exp3}
\end{figure}

The results of experiment 3 are depicted in Fig.~\ref{fig:results_exp3}.
Fig. \ref{fig:results_exp3_no_coupling} shows the temporal course of the global position of both robots that do not perform any coupling.
In this no coupling scenario, robot 2 is driving to both goal points $A$ and $B$ whereas robot 1 only drives to $B$.
Fig.~ \ref{fig:results_exp3_with_coupling} shows the scenario with physical coupling and in-motion transfer.
Here, the robots are coupled for 7 seconds before they uncouple.
Within that time, in-motion transfer can be carried out, before they uncouple.
So, robot 2 can transfer its package with destination $B$ to robot 1 which is why robot 2 drives only to $A$ and robot 1 only to $B$.
A comparison of the navigation metrics of experiment 3 is depicted in Tab. \ref{tab:results_comparison_scenario}.
Our approach enables physical coupling and thus in-motion transfer, which outperforms a no coupling scenario in terms of total time, total energy, and total distance in a simple scenario.
Our approach improves total time by 19.75 \%, total energy by 21.04 \% and total distance by 15.52 \% compared to a no coupling scenario in experiment 3.

\begin{table}[]
    \centering
    \caption{Experiment 3 - Aggregated navigation metrics}
    \label{tab:results_comparison_scenario}
    \begin{tabular}{|c|c|c|c|}
        \hline
                       Approach                                           & Time (s) & Energy (J) & Distance (m) \\ \hline \hline
        No coupling                                                      & 20.25      & 10.49        & 30.2           \\ 
        Ours: with coupling & \textbf{16.25}      & \textbf{8.28}         & \textbf{25.51}          \\ \hline
    \end{tabular}%
 
\end{table}

\subsection{Discussion}\label{subsec:Discussion}
The results show that the approach is capable of physically coupling two omnidirectional wheeled robots in motion.
Moreover, the approach can handle initial states where the docking axes are not perfectly aligned.
This can be seen with the aid of Fig. \ref{fig:results_exp2_slip_docking} and \ref{fig:results_exp2_slip_distance}.
Robot 2 has to be in the approach corridor before the relative distance can go to zero.
That is why Fig. \ref{fig:results_exp2_slip_distance} has a plateau after 2 seconds but decreases after 3.75 seconds, because it is in the docking corridor then.
The soft-docking constraint is the only constraint that increases during the optimization.
But, that has to be the case so that the distance decreases, because a high slip in the soft-docking means driving towards each other.
The simple logistic scenario demonstrates energy and time efficiency and validates that physical coupling can increase efficiency in a logistic scenario.

However, the approach has only been tested in a simulation.
Real-world experiments should be conducted to analyze the performance of our approach with measurements.
More experiments could be done with other system dynamics like differential drive or ackermann kinematic, to evaluate the feasibility for other system dynamics.
\section{CONCLUSIONS}\label{sec:Conclusion}
In this paper, a central \acrshort{mpc} for in-motion physical coupling of mobile ground robots has been designed and implemented in a simulation.
Thereby, we derived an approach for in-motion physical coupling for omnidirectional wheeled robots.
The results validate our approach and demonstrate that in-motion physical coupling of robots is preferable to no coupling in a logistic scenario, as it increases efficiency through in-motion transfer.
Consequently, our approach can be applied to similar modular robots in a logistic context to enhance efficiency by providing the baseline for in-motion transfer and sorting.
Moreover, this approach can be transferred to the mobility sector to enhance efficiency of autnonomous modular pods.
Further work incorporates an evaluation in real world.


\addtolength{\textheight}{-1cm}


\bibliographystyle{IEEEtran}
\bibliography{reference/ITSC_2025}


\end{document}